\documentclass[aps,pre,reprint,groupedaddress]{revtex4-1}
\usepackage{graphicx}
\usepackage{dcolumn}
\usepackage{bm}
\usepackage{epsfig}
\usepackage{subfigure}
\usepackage{graphics}
\usepackage{epstopdf}
\usepackage[english]{babel}
\usepackage{amsfonts}
\usepackage{mathrsfs}
\usepackage{amsthm}
\usepackage{amssymb}
\usepackage{enumerate}
\usepackage{amsmath}
\usepackage{subfigure}
\usepackage{algorithm}
\usepackage{algorithmicx}
\usepackage{algpseudocode}
\usepackage{hyperref}
\usepackage{multirow}
\floatname{algorithm}{Algorithm}

\begin{document}


\title{Core Influence Mechanism on Vertex-Cover Problem through Leaf-Removal-Core Breaking}


\author{Xiangnan Feng$^{1,2}$}
\author{Wei Wei$^{1,2,3}$}
\email[]{weiw@buaa.edu.cn}
\author{Xing Li$^{1,2}$}
\author{Zhiming Zheng$^{1,2,3,*}$}

\affiliation{$^{1}$School of Mathematics and Systems Science, Beihang University, Beijing, China\\
$^{2}$Key Laboratory of Mathematics Informatics Behavioral Semantics, Ministry of Education, China\\
$^{3}$Beijing Advanced Innovation Center for Big Data and Brain Computing, Beihang University, Beijing, China}


\date{\today}
\begin{abstract}
Leaf-Removal process has been widely researched and applied in many mathematical and physical fields to help understand the complex systems, and a lot of problems including the minimal vertex-cover are deeply related to this process and the Leaf-Removal cores. In this paper, based on the structural features of the Leaf-Removal cores, a method named Core Influence is proposed to break the graphs into No-Leaf-Removal-Core ones, which takes advantages of identifying some significant nodes by localized and greedy strategy. By decomposing the minimal vertex-cover problem into the Leaf-Removal cores breaking process and maximal matching of the remained graphs, it is proved that any minimal vertex-covers of the whole graph can be located into these two processes, of which the latter one is a P problem, and the best boundary is achieved at the transition point. Compared with other node importance indices, the Core Influence method could break down the Leaf-Removal cores much faster and get the no-core graphs by removing fewer nodes from the graphs. Also, the vertex-cover numbers resulted from this method are lower than existing node importance measurements, and compared with the exact minimal vertex-cover numbers, this method performs appropriate accuracy and stability at different scales. This research provides a new localized greedy strategy to break the hard Leaf-Removal Cores efficiently and heuristic methods could be constructed to help understand some NP problems.
\end{abstract}


\maketitle

\section{Introduction}

The minimal vertex-cover problem (Vertex-cover) belongs
to one of Karp¡¯s 21 NP-complete problems \cite{Karp1972ibm} and the six basic
NP-complete problems \cite{Cook1971acm, Papadimitriou2003computational}, which has a wide range of applications in
the related real networks, such as immunization strategies in
networks \cite{gardenes2006eur} and monitoring of internet traffic \cite{breitbart2001ieee}.
When the average degree of the graph gets larger than the Euler number
$e$ \cite{weigt2000phys}, the instances of minimum vertex-cover problem on the Erd\"{o}s-R\'{e}nyi random graph
get into the hard-solving region, which is caused by the Leaf-Removal core.
This hard core brings obstacle for solving the minimum vertex-cover, as many correlations
among the nodes/variables produce frustrations (even long-range frustrations \cite{zhou2005phys}) in finding the
optimal solutions, and how to decouple the Leaf-Removal core determines the effect of the solving
strategy.

Generally speaking, Leaf-Removal algorithm plays important roles in understanding complex systems or graphs,
such as detecting the hierarchical architecture and simplification of solving NP problems.
It removes the leaves recursively until no leaves exist,
and the \emph{Leaf-Removal core} is got for the residual graph if it is not a null graph.
Different Leaf-Removal strategies focus on different definitions of the \emph{leaves}:
for giant connected components, the leaves are degree-one nodes with corresponding edges,
and random graphs with average degree larger than 1 have Leaf-Removal core
in the giant connected components in the viewpoint of the Leaf-Removal mechanism \cite{bollobas1985random};
for $k$-XORSAT problem, the leaves are equations with variables only emerging once in the whole problem,
and the Leaf-Removal core can result in the clustering of the all the solutions and one-step replica symmetric
breaking with positive structural entropy \cite{mezard2003alternative};
for minimum vertex-cover problem, the leaves are degree-one nodes with their
neighbors and related edges, and the existence of the Leaf-Removal core produces higher-order
replica symmetric breaking and sets obstruction of understanding the solving complexity
of NP-complete problems \cite{weigt2001physe}. Also, the Leaf-Removal strategy works as an important tool in so many such applications,
such as solving the $k$-SAT problem \cite{Mertens2005Threshold} and \emph{MAS} problem \cite{wei2009jstat}, the $k$-core \cite{dorogovtsev2006kcore} and $k$-shell \cite{kitsak2010identification} organization
of complex network.

The complexity of the Leaf-Removal core can be measured by the cost to break it,
but to break the Leaf-Removal core with the lowest cost (e.g., least number of nodes/edges) always
involves with NP problems, which makes the understanding of its structure complicated.
The collective influence \cite{morone2015influence} was proposed to break the giant connected component, which can also
be viewed as breaking the Leaf-Removal core in the Leaf-Removal viewpoint, and it was studied to approximate
the optimal percolation and reach appropriate accuracy.
The existing algorithms for solving minimum vertex-cover mainly focus on greedy searching strategy
and heuristic strategy. In statistical mechanics, some algorithms based on the solution space structures
were studied and good approximations were achieved in a heuristic way, such as the survey propagation algorithm \cite{weigt2006message} and
the MBEA algorithm \cite{wei2012detecting}. These algorithms took advantage of the backbones and clustering of the solutions
to improve the solving efficiency. In this paper, it is aimed to break the Leaf-Removal
core of the minimum vertex-cover problem in a greedy way, an order for all the nodes will be investigated to
break the Leaf-Removal core as fast as possible, and combined
with some easily coverage method for the residual graph,
which can be used to obtain the approximated optimal solutions.


\section{Leaf-Removal for minimum vertex-cover}
\subsection{The Leaf-Removal Process}
As mentioned before, there are different definitions on the leaves in the graphs. In this paper, the leaves in the graph stand for the nodes of which the degrees are one and their neighbors, namely, removing the leaves means to remove the one-degree nodes and their neighbors. At the same time, all the links connected to these nodes are removed from the graph.
\begin{figure}[htbp]
	\includegraphics[scale=0.82]{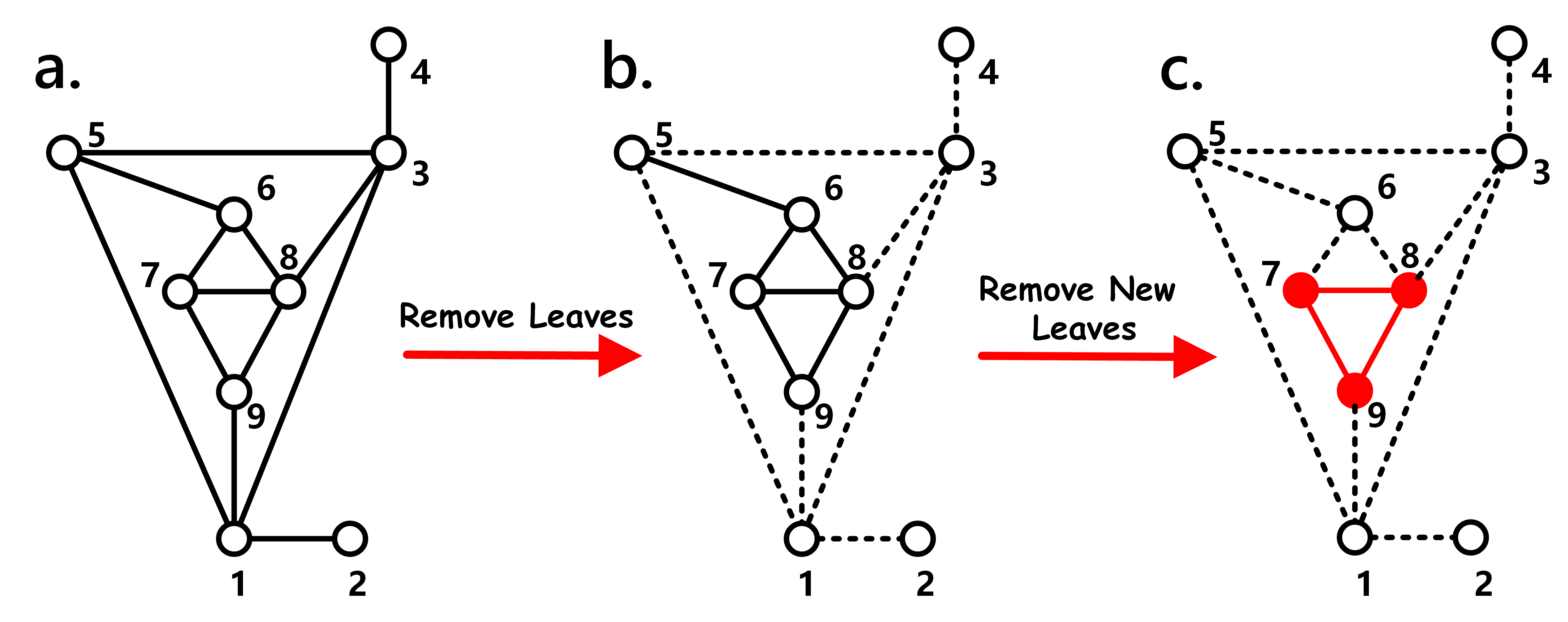}
	\caption{The Leaf-Removal process of the graph in \textbf{a}. $v_1\sim v_2$ and $v_3\sim v_4$ are the two leaves. The graph in \textbf{b} is got with the two leaves removed, and in the remained graph, $v_5\sim v_6$ is a new leaf. Removing this new leaf, we could get the graph in \textbf{c}. There is no more one-degree node and the remained triangle composed by $v_7$, $v_8$ and $v_9$ is the Leaf-Removal core, which is marked as red in \textbf{c}. The dotted lines denote the deleted edges when nodes are removed.}\label{Leaf-Removal}
\end{figure}

Usually, when leaves are removed from the graph, new leaves will appear and removing the new leaves may bring more new leaves. The Leaf-Removal is a recursive process. Keep removing leaves in the new graph until there is no leaves in the remained graphs. Clearly the remained subgraph is composed of some isolated nodes and several relatively densely connected cores, the \emph{Leaf-Removal cores}. There are no nodes in the cores whose degree is one. An example is shown in Figure~\ref{Leaf-Removal} to perform a complete Leaf-Removal process and identify the Leaf-Removal core.
\subsection{Greedy strategy for the coverage by breaking the Leaf-Removal core}

For an undirected graph $G = (V,E)$ containing $N$ nodes and $M$ edges where $V$ is the vertex set and $E$ is the edges set, let $C$ be a set of specific nodes in graph $G$. $C(G)$ is called to be a cover of $G$ if for every edge $(v_i,v_j)\in E$, there is $v_i\in C(G)$ or $v_j\in C(G)$. The minimum vertex-cover, denoted as $C_m(G)$, is defined as the minimal of $C(G)$:
\begin{eqnarray}
C_m(G) = \mathop{\arg\min}_C\  |C(G)|.
\end{eqnarray}

Let $\textbf{n} = (n_1,n_2,\dots, n_N)$ be the indices array of graph $G$ where $n_i = 0$ indicates that node $v_i$ and all the edges connected to it are removed from the graph and otherwise $n_i = 1$. In this way the graph induced by the remaining nodes $i$ with $n_i=1$ is denoted as $G(\textbf{n})$, and the corresponding vertex-cover $C(G)$ and minimum vertex-cover $C_m(G)$ are changed into $C(G(\textbf{n}))$ and $C_m(G(\textbf{n}))$. The number of remained nodes is $N(\textbf{n}) = \sum_{i = 1}^{N}n_i$ and the vertex set is $V(\textbf{n}) = \{v_i\in V|n_i=1\}$.

By deleting nodes one by one from graph $G$, it is trivial to see that at a certain step the graph will be turned into a No-Leaf-Removal-Core graph. Assume following a specific method, the nodes are deleted in a certain order, and the nodes indices array are $\textbf{n}_0=(1,1,\dots,1)$, $\textbf{n}_1$, $\textbf{n}_2$, $\dots$, $\textbf{n}_N=(0,0,\dots,0)$. At a certain step $t$, the graph turns into a no-core graph, namely $G(\textbf{n}_t)$ contains no Leaf-Removal cores. $G(\textbf{n}_i)$ is not a no-core graph for all $i< t$ and for all $t\leq i\leq N$,  $G(\textbf{n}_i)$ is no-core graph. The $t$ is the transition point.

\begin{figure}[htbp]
	\includegraphics[scale=0.62]{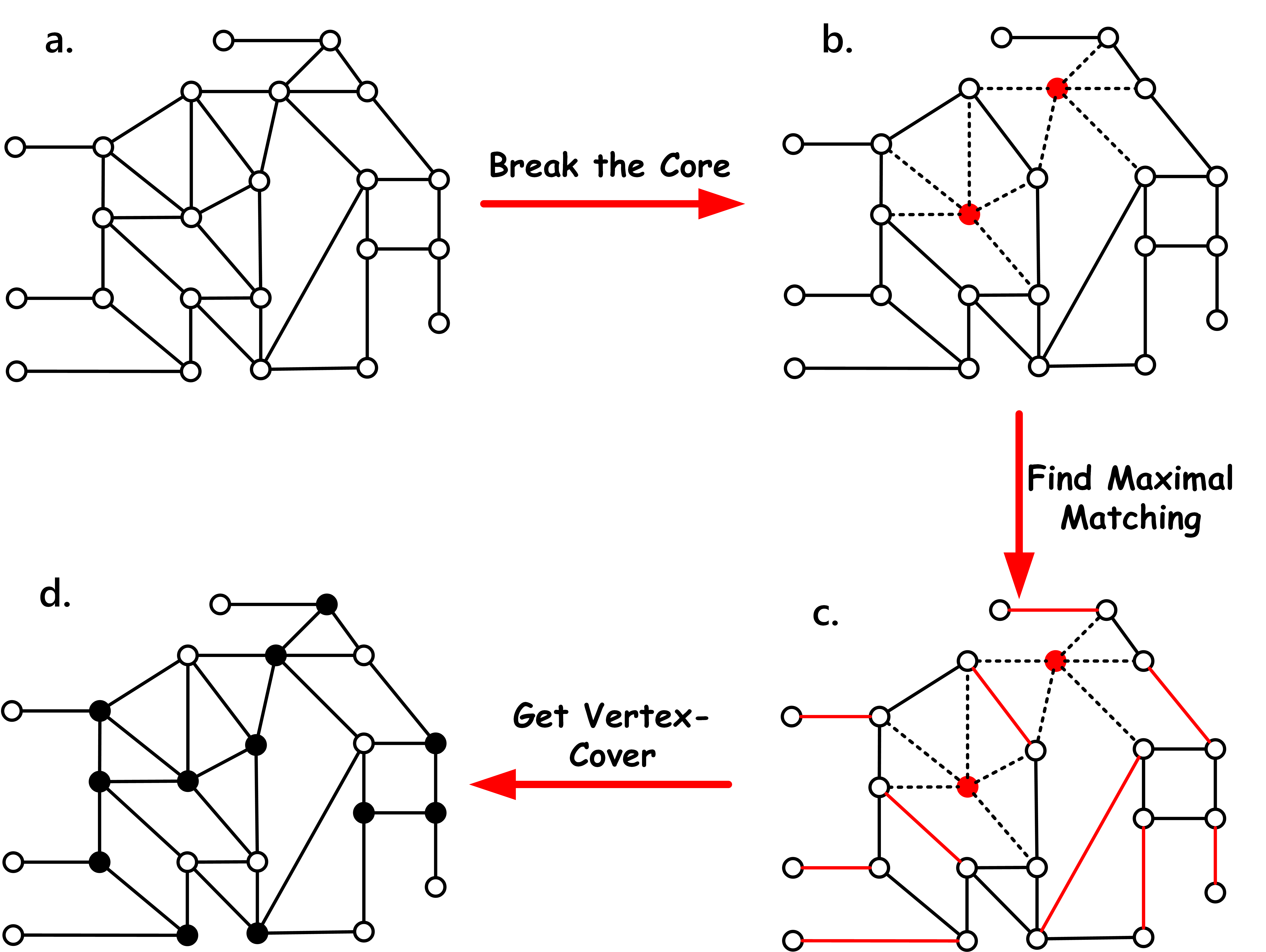}
	\caption{The procedures to get a vertex-cover of graph in \textbf{a} by the strategy of deleting the graph into a no-core one. \textbf{a} owns one Leaf-Removal core. By the algorithm proposed in the following section, the red nodes in \textbf{b} are removed and the remained graph is a no-core graph. The remained graph satisfies the property that the maximal matching number equals to the minimal cover number. In \textbf{c}, one maximal match is marked by red edges. Combing the deleting nodes in \textbf{b} and results of the maximal matching in \textbf{c}, the vertex-cover could be got, denoted by black nodes in \textbf{d}. The dotted lines denote the deleted links when nodes are removed. The maximal matching number in \textbf{c} is $10$ and thus the cover number of graph in \textbf{a} got by this strategy is $12$, which exactly equals to the minimal cover number of this graph.}\label{example-graph}
\end{figure}

By the K\"{o}nig's theorem, the residual subgraph
$G(\textbf{n}_j)$ for $t\leq j\leq N$ has no Leaf-Removal core and is a K\"{o}nig-Egerv\'{a}ry graph.
Its minimum coverage number is equal to its maximal matching number $M(G(\textbf{n}_j))$ \cite{bondy1976graph},
and finding the maximal matching for a graph is an easily-solving problem.
If all the removed nodes $v_i$ (namely with $n_i=0$ in $\textbf{n}_j$ where $t\leq j\leq N$) are covered,
a greedy strategy for approximation of the minimal coverage can be achieved:
for any vertex-cover $C(G)$ of graph $G$,
there exists a deleting order or method which produces node arrays
$\textbf{n}_0$, $\textbf{n}_1$, $\dots$, $\textbf{n}_N$ with transition point $t$,
at the step $t$ the cover of the graph $G$ could be decomposed as:
\begin{eqnarray}
C(G) = (V\setminus V(\textbf{n}_t))\cup C_m(G(\textbf{n}_t)),
\end{eqnarray}
and the cover number is:
\begin{eqnarray}
|C(G)| = t+ M(G(\textbf{n}_t)).
\end{eqnarray}
An example about getting a cover of a graph through this strategy which deletes the graph into a no-core one is presented in Figure~\ref{example-graph}. The maximum matching can be solved by an algorithm within polynomial time for general graphs \cite{gabow1991faster}, and the main effort for this greedy strategy relies on determining an optimal
deleting order to make the transition point $t$ as small as possible.
The proposed strategy aims to break the Leaf-Removal core with the lowest cost,
and how to distinguish the important vertex and recognize the polarized vertices
as the cover backbones is the core difficulty.

\subsection{Relations between the core-breaking and optimal solutions}

\textbf{\emph{Proposition:}} For nodes arrays $\textbf{n}_i$ and $\textbf{n}_j$
where $i\geq j\geq t$, the coverage numbers for the whole graph $G$ by
the two core-breaking ways satisfy $|C(G(\textbf{n}_i))+i|\geq |C(G(\textbf{n}_j))+j|$. \\
\emph{Proof}:
When one more node is deleted from $G(\textbf{n}_i)$ and the graph becomes $G(\textbf{n}_{i+1})$
with $N(\textbf{n}_{i+1}) +1= N(\textbf{n}_{i})$.
Yet since one node owns to at most one match,
the maximal matching number of $G(\textbf{n}_{i+1})$ will decrease by one or remain the same to $G(\textbf{n}_{i})$,
namely $M(G(\textbf{n}_{i+1})) \geq M(G(\textbf{n}_{i})) -1$.
In this way $|C(G(\textbf{n}_{i+1}))+i+1|\leq|C(G(\textbf{n}_i))+i|$ and
$|C(G(\textbf{n}_i))+i|\geq |C(G(\textbf{n}_j))+j|$ for all $i\geq j\geq t$. $\Box$

This proposition shows that the best approximation for the minimal coverage under the above
core-breaking strategy occurs at the transition point $t$, i.e., the first time that Leaf-Removal
cores disappear, and so our algorithm greedily deletes the least number of nodes to produce a
no-core subgraph with easy minimal coverage. It is evident that the obtained coverage
is not necessary to be a minimal one, but the following theorem illustrates that all the minimal
vertex-cover can be found in such a core-breaking way.

\textbf{\emph{Theorem:}}
For any graph $G$ and its minimal vertex-cover $C_m(G)$, there exists a deleting method to produce node arrays
$\textbf{n}_0$, $\textbf{n}_1$, $\dots$, $\textbf{n}_N$ and transition point $t$,
such that this minimal vertex-cover could be represented as:
\begin{eqnarray}
\label{mvc1}
C_m(G) = (V\setminus V(\textbf{n}_t))\cup C_m(G(\textbf{n}_t)),
\end{eqnarray}
and the minimal cover number is:
\begin{eqnarray}
\label{mvc2}
|C_m(G)| = t + M(G(\textbf{n}_t)).
\end{eqnarray}\\
\emph{Proof}:
Without generality, the minimal vertex-cover has $v_1, \cdots, v_s$ as covered nodes with
$v_{s+1}, \cdots, v_N$ as uncovered nodes of graph $G$. 
For any core-breaking strategy $\textbf{n}_0$, $\textbf{n}_1$, $\dots$, $\textbf{n}_t$,
we have $t<s$. And, if we delete the nodes according to the natural order
$v_1, \cdots, v_s$, i.e., ensuring all the deleted nodes are covered one,
after the core-breaking strategy $\textbf{n}_0$, $\textbf{n}_1$, $\dots$, $\textbf{n}_s$,
the residual graph $G(\textbf{n}_s)$ should be a null graph composed by isolated nodes. The original
graph has Leaf-Removal cores and residual graph $G(\textbf{n}_s)$ has no core,
then there must exists a certain number $t$ that the core is firstly broken, and the minimal
vertex-covered nodes evidently have a decomposition representation 
$C_m(G) = (V\setminus V(\textbf{n}_t))\cup C_m(G(\textbf{n}_t))$ with 
$|C_m(G)| = t + M(G(\textbf{n}_t))$.
$\Box$

Usually for one minimal vertex-cover of a graph,
there are more than one corresponding deleting methods. 
The deleting order of the nodes by the core-breaking order can
also be arranged in other greedy strategy, such as deleting
the covered nodes with biggest degrees. As for a no-core graph, each leaf
contains a covered node with the other uncovered, and the number of covered nodes 
must be fewer or equal to the number of uncovered ones,
thus generally after deleting at least $max\{0,N-(N-s)*2\}=max\{0,2s-N\}$
covered nodes, it will produce a no-core graph which satisfies $t\geq max\{0,2s-N\}$. Besides, by deleting
one of neighbored covered nodes for all such pairs, the residual graph has 
each covered node with no covered neighbors, and it is almost a no-core graph
with high probability; when the original graph has no triangles, this deleting
strategy only deletes at most $s/2$ nodes and $t\leq s/2$ 
with high probability to break the Leaf-Removal core.
There are many other strategies to break the Leaf-Removal core, such as 
deleting the nodes in the order of different centralities or node importance measurements.

When a graph has Leaf-Removal cores, with a high probability 
it is not a K\"{o}nig-Egerv\'{a}ry subgraph 
and the relation between the maximal match, the minimal vertex-cover and its vertex number is quite complicated. 
Though a minimal vertex-cover can be found in such a core-breaking way, but the fast way
to break the Leaf-Removal core does not always corresponds to achieve a minimal vertex-cover.
As the problem of breaking the giant connected component is NP hard,
the optimal solution of breaking the Leaf-Removal cores is also a hard problem,
which should be related to but different from the minimal vertex-cover problem.
Here, it is safe to say that the minimal vertex-cover does not always 
correspond to the fastest method to delete the graph into a no-core graph. 
Yet, it is meaningful to study the process of deleting nodes since 
it provides us a new perspective to understand the complicated 
organization of the Leaf-Removal core and study the minimal vertex-cover problem.

\section{Delete a Graph into the No-Core Graph}
In this section we will discuss how to turn graphs into no-core ones by deleting nodes with the lowest cost. As we said before, this problem will concern a lot of applications in real-world networks.

The Leaf-Removal cores will appear after the leaf-removing process is completed. For node $v_i$, we define $c_i$ in $\textbf{c}=(c_1,c_2,\dots,c_N)$ to indicate the probability of node $v_i$ belonging to the final remained core:
\begin{eqnarray}
c_i=\left\{
\begin{array}{ll}
1 & \textrm{if $v_i$ is in a Leaf-Removal core}\\
0 & \textrm{otherwise}.
\end{array} \right.
\end{eqnarray}
Obviously, if the degree of node $v_i$ is one or node $v_i$ is connected to a one-degree node in the Leaf-Removal process, we have $c_i=0$.

The problem of breaking the Leaf-Removal cores could be described in the language of optimization. For a node belonging to the original graph, to delete the graph into a no-core one, either this node is removed, or this node does not belong to a Leaf-Removal core. Thus, this problem could be regarded as find the $\mathbf{n} = (n_1,n_2,\dots,n_N)$ satisfying:
\begin{eqnarray}
\textrm{$\mathbf{n} = \mathop{\arg\min}_{\mathbf{n}}\  N(\mathbf{n})$,
subject to $\sum_{i=1}^{N}c_i n_i =0$}.
\end{eqnarray}

Consider two connected nodes $v_i$ and $v_j$ in the original graphs. It is clear that removing node $v_j$ may effect the value of $c_i$, yet this relation is complex. There are four situations for $c_i$ and $c_j$: $c_i=0$ and $c_j=0$; $c_i=0$ and $c_j=1$; $c_i=1$ and $c_j=0$; $c_i=1$ and $c_j=1$.

We will ignore the first two situations where node $v_i$ and node $v_j$ do not belong to the cores because in our following proposed model this situation will not effect the results. In the third situation, the node $v_j$ will be removed in one of the leaves removing steps. Clearly, the degree of node $v_j$ will not be one and it is connected to a one-degree node, because if $v_j$ is a one-degree node, node $v_i$ will be deleted. Thus the removal of node $v_j$ will not effect the value of $c_i$. The forth situation is complicated: when the core is densely connected, the removal of $v_j$ may have no effect on $c_i$; if the core is not that dense or uneven, removing of node $v_j$ will produce new leaves and more nodes will be removed in the original cores.

Let $v_{i\to j}$ be the probability of $v_i$ belonging to the core with $v_j$ removed. Consider all the nodes that are connected to $v_i$. The value of $c_i$ with $v_j$ removed from the graph is determined by the status of all the neighbors of $v_i$. For a network with tree-like structure locally, this relation could be formulated by:
\begin{eqnarray}
c_{i\to j} = c_i n_i[1-\prod_{h\in \partial i, h\neq j}(1-c_{h\to i})],
\end{eqnarray}
where the $\partial i$ is the set of neighbors of node $v_i$. This system takes into account the first two situations we have mentioned above: if node $v_i$ does not belong to any of the final Leaf-Removal cores, then $c_{i\to j}$ will keep to be $0$ no matter whether node $v_j$ is removed or not. Obvious $c_{i\to j}=0$ if $n_i=0$, namely node $v_i$ is removed from the graph.

In this system, the set $\{c_{i\to j}=0|\forall v_i,v_j\in V\}$ will always be one solution. It is worth noting that this case will not always correspond to a no-core graph. For example, a weakly connected network, like a cycle, could be very fragile and removing any node will turn the graph into a no-core one, although in the original graph all the nodes belong to the Leaf-Removal core. Yet it is clear that this kind of graph is very unstable and the stable ones are those whose connection is much denser. It is pretty hard to get the unstable Leaf-Removal cores like the cycle by Leaf-Removal process since only one leaf could easily break it. So in this paper we will neglect this situation and regard that the solution $\{c_{i\to j}=0|\forall v_i,v_j\in V\}$ corresponds to an no-core graph.

With knowledge in dynamic system, the solution will be stable if the largest eigenvalue of the linear operation $\mathop{R}$ is smaller than one \cite{wiggins2003introduction}. $\mathop{R}$ is a matrix with $2M$ rows and $2M$ columns and each row or column stands for an edge with one direction $i\to j$. Since edge between $v_i$ and $v_j$ has two directions $i\to j$ and $j\to i$, this matrix will take into account all the directions and defined as:
\begin{eqnarray}
\mathop{R_{w\to x, y\to z}}=\frac{\partial c_{w\to x}}{\partial c_{y\to z}}|_{c_{w\to x=0}}.
\end{eqnarray}
Let $\lambda(\textbf{c},\textbf{n})$ be the largest eigenvalue of matrix $\mathop{R}$ and clearly this value is determined by index $\textbf{n}$ and $\textbf{c}$. The stability of a solution corresponding to no-core graph is determined by $\lambda(\textbf{c},\textbf{n})<1$.

It could be found that, in a graph with tree like structure locally, this matrix $\mathop{R}$ is calculated by the non-backtracking matrix \cite{bordenave2015non, hashimoto1989in} $\mathop{B}$:
\begin{eqnarray}
\mathop{R_{w\to x, y\to z}} = c_y n_y \mathop{B_{w\to x, y\to z}},
\end{eqnarray}
where $\mathop{B}$ is defined as:
\begin{eqnarray}
\mathop{B_{w\to x, y\to z}} = \left\{
\begin{array}{ll}
1 & \textrm{if $x=y$, $w\neq z$}\\
0 & \textrm{otherwise}.
\end{array} \right.
\end{eqnarray}

The non-backtracking has received a lot of attention recently and been used to identify the non-backtracking walks on graphs. According to the Perron-Frobenius theorem \cite{horn2012matrix}, since all the entries in $\mathop{B}$ is non-negative, it largest eigenvalue and all the entries of the corresponding eigenvector are all positive. Krzakala et al. showed that the spectrul method based on the non-backtracking method, namely the eigenvector corresponding to the second largest, could be applied to divide communities in graphs and could reach the detectability threshold \cite{krzakala2013spectral}. Neumann et al. used this matrix to overcome the pathology of nodes near the high-degree nodes or hubs when eigenvector centrality is applied in some graphs to evaluate the influence of nodes \cite{percolation2014newman}. Hernan et al. applied the non-backtracking matrix to the optimal percolation problem and find an index to reduce the size of giant component \cite{albert2000error, cohen2001breakdown}, which is crucial in influence optimization and immunization \cite{morone2015influence}.

As nodes get deleted in graphs, the largest eigenvalue $\lambda({\textbf{c},\textbf{n}})$ of the modified non-backtracking matrix $\mathop{R}$ changes. When a node is deleted or the the node is no longer in the cores, the corresponding entries in $\mathop{R}$ turn into zero. According to the Perron-Frobenius theorem, the largest eigenvalue will decrease. The problem of deleting the graph into a no-core one is then equal to find the fastest way to reduce the largest eigenvalue to be less than one, namely this problem could be represented as:
\begin{eqnarray}
\textrm{$\min |N-N(\textbf{n})|$, subject to $\lambda({\textbf{n},\textbf{c}})<1$}.
\end{eqnarray}
It is impossible to solve it directly, since $\textbf{n}$ and $\textbf{c}$ is sophisticatedly tangled and there is no formula of $\lambda({\textbf{n}, \textbf{c}})$ in the form of $\textbf{n}$ and $\textbf{c}$. To overcome this, we will apply the power method \cite{kincaid2009numerical}, a classical numerical technique, to approximate the largest eigenvalue.

Let $\textbf{v}_0$ be a vector with nonzero projection on the direction of the eigenvector corresponding to $\lambda(\textbf{n},\textbf{c})$ and $\textbf{v}_l(\textbf{n},\textbf{c})$ be the vector after $l$ times iterations by $\mathop{R}$:
\begin{eqnarray}
\textbf{v}_l(\textbf{n},\textbf{c})=\mathop{R^l}\textbf{v}_0.
\end{eqnarray}
By power method, we have:
\begin{eqnarray}
\lambda(\textbf{n},\textbf{c})=\lim_{l\to \infty}\lambda_l(\textbf{n},\textbf{c})=\lim_{l\to \infty}\bigl(\frac{|\textbf{v}_l(\textbf{n},\textbf{c})|}{|\textbf{v}_0|}\bigr)^{\frac{1}{l}},
\end{eqnarray}
where for $\textbf{v}_l(\textbf{n},\textbf{c})$, we have:
\begin{eqnarray}
|\textbf{v}_l(\textbf{n},\textbf{c})|^2 && = \langle \textbf{v}_l(\textbf{n},\textbf{c})|\textbf{v}_l(\textbf{n},\textbf{c})\rangle\\
&&=\langle \textbf{v}_0|(\mathop{R^l})^T\mathop{R^l}|\textbf{v}_0\rangle.
\end{eqnarray}

Firstly, for $l=1$, the approximated eigenvector corresponding to $\lambda(\textbf{n},\textbf{c})$ is:
\begin{eqnarray}
|\textbf{v}_1(\textbf{n},\textbf{c})\rangle = \mathop{R}|\textbf{v}_0\rangle.
\end{eqnarray}
To further calculate this, the matrix $\mathop{R}$ could be calculated by:
\begin{eqnarray}
\mathop{R_{w\to x, y\to z}} = c_y n_yA_{wx}A_{yz}\delta_{xy}(1-\delta_{wz}),
\end{eqnarray}
where the $A_{wx}$ and $A_{yz}$ are entries of the adjacency matrix and $\delta_{xy}$ and $\delta_{wz}$ are the Kronecker symbols with:
\begin{eqnarray}
\delta_{ij}=\left\{
\begin{array}{ll}
1 & \textrm{if $i=j$}\\
0 & \textrm{otherwise}.
\end{array} \right.
\end{eqnarray}
The $n_y$ and $c_y$ guarantee the node $v_y$ is not deleted from the graph and belongs to the core. The $A_{wx}$, $A_{yz}$ and $\delta_{xy}$ ensure that there is a track from $v_w$ to $v_z$ through $v_x$ or $v_y$ and the $\delta_{wz}$ ensure the non-backtracking property.

For simplicity, let $s_i$ stand for the state of node $v_i$ and $q_i$ the excess degree namely:
\begin{eqnarray}
s_i \equiv c_in_i, \quad q_i = d_i-1.
\end{eqnarray}
Choosing the $2M$-dimension vector $|\textbf{v}_0\rangle = |\textbf{1}\rangle$, we could get the left vector:
\begin{eqnarray}
_{w\to x}\langle\textbf{v}_1(\textbf{n},\textbf{c})| && = \sum_{y\to z}{_{y\to z}}\langle\textbf{v}_0|\mathop{R_{w\to x,y\to z}}\nonumber\\
&& = s_wA_{wx}q_w,
\end{eqnarray}
and the right vector:
\begin{eqnarray}
|\textbf{v}_1(\textbf{n},\textbf{c})\rangle_{w\to x} && = \sum_{y\to z}\mathop{R_{w\to x,y\to z}}|\textbf{v}_0\rangle_{y\to z}\nonumber\\
&& = s_x A_{wx} q_x.
\end{eqnarray}
Thus, the norm could be presented as:
\begin{eqnarray}
|\textbf{v}_1(\textbf{n},\textbf{c})|^2 && = \sum_{w\to x}{_{w\to x}}\langle\textbf{v}_1(\textbf{n},\textbf{c}) | \textbf{v}_1(\textbf{n},\textbf{c})\rangle_{w\to x}\\
&& = \sum_{w\to x}A_{wx}q_w q_x s_w s_x.\nonumber
\end{eqnarray}
Since $|\textbf{v}_0|^2=\sum_{i=1}^{N}d_i = 2M$, we have:
\begin{eqnarray}
\lambda_1(\textbf{n},\textbf{c}) = \bigl(\frac{1}{2M}\sum_{w\to x}A_{wx}q_w q_x s_w s_x\bigr)^{\frac{1}{2}}.
\end{eqnarray}

From the equation above, it could be observed that the approximation to the largest eigenvalue at order $1$ $\lambda_1(\textbf{n},\textbf{c})$ involve all the edges in the graphs, or in other word, all the paths of which length is one. The non-backtracking requires that if there exist self-edges, they should not be taken into account in the approximation.

For $l=2$, the left vector $_{w\to x}\langle\textbf{v}_2(\textbf{n},\textbf{c})|$ could be derived directly from the same process:
\begin{eqnarray}
_{w\to x}\langle\textbf{v}_2(\textbf{n},\textbf{c})| && = \sum_{y\to z}{_{y\to z}}\langle\textbf{v}_1|\mathop{R_{w\to x,y\to z}}\nonumber\\
&& = s_wA_{wx}\sum_{i}^{N} A_{wi} s_i q_i (1-\delta_{xi}),
\end{eqnarray}
and the same to the right vector:
\begin{eqnarray}
|\textbf{v}_2(\textbf{n},\textbf{c})\rangle_{w\to x} && = \sum_{y\to z}\mathop{R_{w\to x,y\to z}}|\textbf{v}_1\rangle_{y\to z}\nonumber\\
&& = s_x A_{wx}\sum_{i}^{N} A_{xi}s_i q_i (1-\delta_{wi}).
\end{eqnarray}
So the norm is:
\begin{eqnarray}
|\textbf{v}_2(\textbf{n},\textbf{v})|^2 = && \sum_{w\to x, y\to z}(A_{wx} A_{xy} A_{yz}\times \nonumber\\ && (1-\delta_{wy}) (1-\delta_{xz})q_w q_z s_w s_x s_y s_z),
\end{eqnarray}
and the approximation to the largest eigenvalue $\lambda_2(\textbf{n},\textbf{c})$ is:
\begin{eqnarray}
\lambda_2(\textbf{n},\textbf{c}) = && \bigl(\frac{1}{2M}\sum_{w\to x, y\to z}(A_{wx}A_{xy} A_{yz}\times \nonumber\\ && (1-\delta_{wy}) (1-\delta_{xz})q_w q_z s_w s_x s_y s_z) \bigr)^{\frac{1}{4}}.
\end{eqnarray}

By the same derivation, the approximation at order $3$ to the largest eigenvalue $\lambda_3(\textbf{n},\textbf{c})$ is:
\begin{eqnarray}
\lambda_3(\textbf{n},\textbf{c}) = && \bigl(\frac{1}{2M}\sum_{\begin{subarray}{c}u\to v, w\to x\\ y\to z\end{subarray}}(A_{uv}A_{vw}A_{wx}A_{xy} A_{yz}\times \nonumber\\ && (1-\delta_{uw}) (1-\delta_{vx}) (1-\delta_{wy}) (1-\delta_{xz})\times \nonumber\\
&& q_u q_z s_u s_v s_w s_x s_y s_z) \bigr)^{\frac{1}{6}}.
\end{eqnarray}
And for any $l>0$, we have:
\begin{eqnarray}
\lambda_l(\textbf{n},\textbf{c}) = && \bigl(\frac{1}{2M}\sum_{x_1,x_2,\dots, x_{2l}} (A_{x_1x_2}A_{x_2x_3}\dots A_{x_{2l-1}x_{2l}}\times \nonumber\\
&& (1-\delta_{x_1x_3})(1-\delta_{x_2x_4})\dots(1-\delta_{x_{2l-2}x_{2l}})\times \nonumber \\
&& q_{x_1}q_{x_{2l}}s_{x_1}s_{x_2}\dots s_{x_{2l}})\bigr)^{\frac{1}{2l}}
\end{eqnarray}

As we could see, the $l=2$ approximation involves all the paths of length $3$ and all the pathes satisfy the non-backtracking property; the $l =3$ approximation require all the paths of length $5$ to be searched. It could be inferred that in the order $l$ approximation, the paths whose length is $2l-1$ are needed to be considered. It is worth noting that the in the approximation with order larger than $2$, when a path meets the non-backtracking requirement, loops are allowed in the path. The track could walk by the nodes that have been already visited in the path.

The formula of the approximation to the largest eigenvalue at order $l$ could be written in other way:
\begin{eqnarray}
\lambda_l(\textbf{n},\textbf{c}) = \bigl(\sum_{i = 1}^{N}q_i\sum_{j\in \partial (i,2l-1)}(\prod_{k\in \mathop{P_{2l-1}}(i,j)}s_k)q_j \bigr)^{\frac{1}{2l}},
\end{eqnarray}
where the $\partial (i,2l-1)$ contains all the nodes $v_j$ which has paths of length $2l-1$ (with loops allowed), and $\mathop{P_{2l-1}}(i,j)$ contain all the nodes which are on these paths.

The formula above involves all the paths in the graph with length $2l-1$. Yet this calculation could be generalized to the situation of length $2l$. By Power Method, the largest eigenvalue could also approximated by:
\begin{eqnarray}
\lambda(\textbf{n},\textbf{c})=\lim_{l\to \infty}\bigl(\frac{\langle \textbf{v}_l(\textbf{n},\textbf{c})|\mathop{R}|\textbf{v}_l(\textbf{n},\textbf{c})\rangle}{\langle\textbf{v}_0|\textbf{v}_0\rangle}\bigr)^{\frac{1}{2l+1}}.
\end{eqnarray}
For example, when $l = 1$, we have:
\begin{eqnarray}
\langle \textbf{v}_1(\textbf{n},\textbf{c})|\mathop{R}|\textbf{v}_1(\textbf{n},\textbf{c})\rangle = \sum_{x\to y, y\to z} && A_{xy}A_{yz}(1-\delta_{xz})\nonumber \\
&& \times q_x q_z s_x s_y s_z.
\end{eqnarray}
Thus, for any $l$, according to the same derivation introduced above, the largest eigenvalue could be approximated by:
\begin{eqnarray}
\lambda'_l(\textbf{n},\textbf{c}) = \bigl(\sum_{i = 1}^{N}q_i\sum_{j\in \partial (i,2l)}(\prod_{k\in \mathop{P_{2l}}(i,j)}s_k)q_j \bigr)^{\frac{1}{2l+1}}.
\end{eqnarray}

From all the calculation above, when paths of length $l$ are considered in the approximation, the leading part of the approximation formula is:
\begin{eqnarray}
\mathop{H_l} = \sum_{i = 1}^{N}q_i\sum_{j\in \partial (i,l)}(\prod_{k\in \mathop{P_{l}}(i,j)}s_k)q_j.
\end{eqnarray}
To reduce the largest eigenvalue quickly, a direct idea is to find the nodes which could reduce the $\mathop{H_l}$ fastest. To illustrate this, we define the \emph{Core Influence of node $v_i$}, $H(v_i)$ to be:
\begin{eqnarray}
\label{Hvi}
H(v_i)=q_i\sum_{j\in \partial (i,l)}(\prod_{k\in \mathop{P_{l}}(i,j)}s_k)q_j.
\end{eqnarray}
This value will be the indicator in removing nodes to turn the graph into a no-core one. First calculate $H(v_i)$ for all the nodes and deleted the node who owns the highest $H(v_i)$ value and all the edges connected to this node. Repeat this process again until the remained graphs is a no-core one. The detailed algorithm is presented in~\ref{algo1}. After the graph is turned into a no-core one, find the maximal match for the remained graph. By unifying the deleted nodes and matched nodes, a vertex-cover could be got.
\begin{algorithm}[H]
	\caption{The process of deleting the graph into a no-core one.}
	\begin{algorithmic}
		\State {\bf Input} graph $G$;
        \While{$G$ is not a no-core graph, namely $\sum c_i\neq 0$}
        \State For each node $v_i$, calculate $c_i$, $q_i$ and the state values $s_i=c_in_i$;
		\State For each node $v_i$, calculate all the Core Influence $H(v_i)$;
		\State Select node $v_j$ with largest $H(v_j)$ value;
		\State Remove node $v_j$ and all the edges linked to it from the graph;
		\State Update the graph $G$, set $n_i=0$;
		\EndWhile
		\State {\bf Output} Nodes that are deleted in each step, namely $\mathbf{n}$.
	\end{algorithmic}
\label{algo1}
\end{algorithm}

The calculation complexity of Leaf-Removal process is $\mathop{O}(N)$, and calculating the Core Influence takes $\mathop{O}(N\log N)$ complexity. In this way, to spot and remove the node with the highest Core Influence costs $\mathop{O}(N)$ time complexity. In a graph, around half of the total nodes need to be removed to turn the graph into a no-core one. Thus, the total complexity of Algorithm~\ref{algo1} is $\mathop{O}(N^2\log N)$.

One problem in applying the Algorithm~\ref{algo1} is how to update the state indicator values $c_i$. The leaf-removing is a fast process with time complexity $\mathop{O}(N)$, however, it will cost a lot to update them for every step. To overcome this, we propose to use a relatively coarse yet much faster strategy to update the indicator values. Before node $v_i$ is removed from the graph, we search all the neighbors of $v_i$. Then neighbors with degree two, namely the nodes which will become leaves if $v_i$ is removed, are specified. The state values $s$ of these neighbors and nodes connected to them are set to be zero. A detailed algorithm is shown in~\ref{algo2}. This will reduce the complexity to $\mathop{O}(N(\log N)^2)$. Since the structure of the core is sophisticated, removing one node in core may produce a long-lasting effect on the core structure and more nodes rather than neighbors will turn into leaves (of higher orders) and be deleted from the core, this method in~\ref{algo2} is only a local approximation to the complete update process. Better methods require a much deeper understanding to the core structure, which deserves further research.

\begin{algorithm}[H]
	\caption{The updating method of indicator values.}
	\begin{algorithmic}
		\State {\bf Input} graph $G$, state values $c$ and $n$ for each node, node $v_i$ that is going to be deleted;
        \For {Each neighbor $v_j$ of node $v_i$}
		\If {$d_j ==2$, the two nodes connected to $v_j$ is $v_i$ and $v_k$}
        \State Set $c_j=0$, $c_i=1$, $c_k=1$;
		\EndIf
        \EndFor
	\end{algorithmic}
\label{algo2}
\end{algorithm}

\section{Experiments}
In this section some experiments are conducted to present the efficiency of several centrality methods of deleting nodes. We examine the steps required to delete the graphs into no-core ones by each centrality and compare the numbers of steps. Also, combining with the theorems mentioned above, the minimal cover numbers of all the methods are compared.
\subsection{Results for Leaf-Removal Cores Breaking}
\begin{figure}[htbp]
	\includegraphics[scale=0.07]{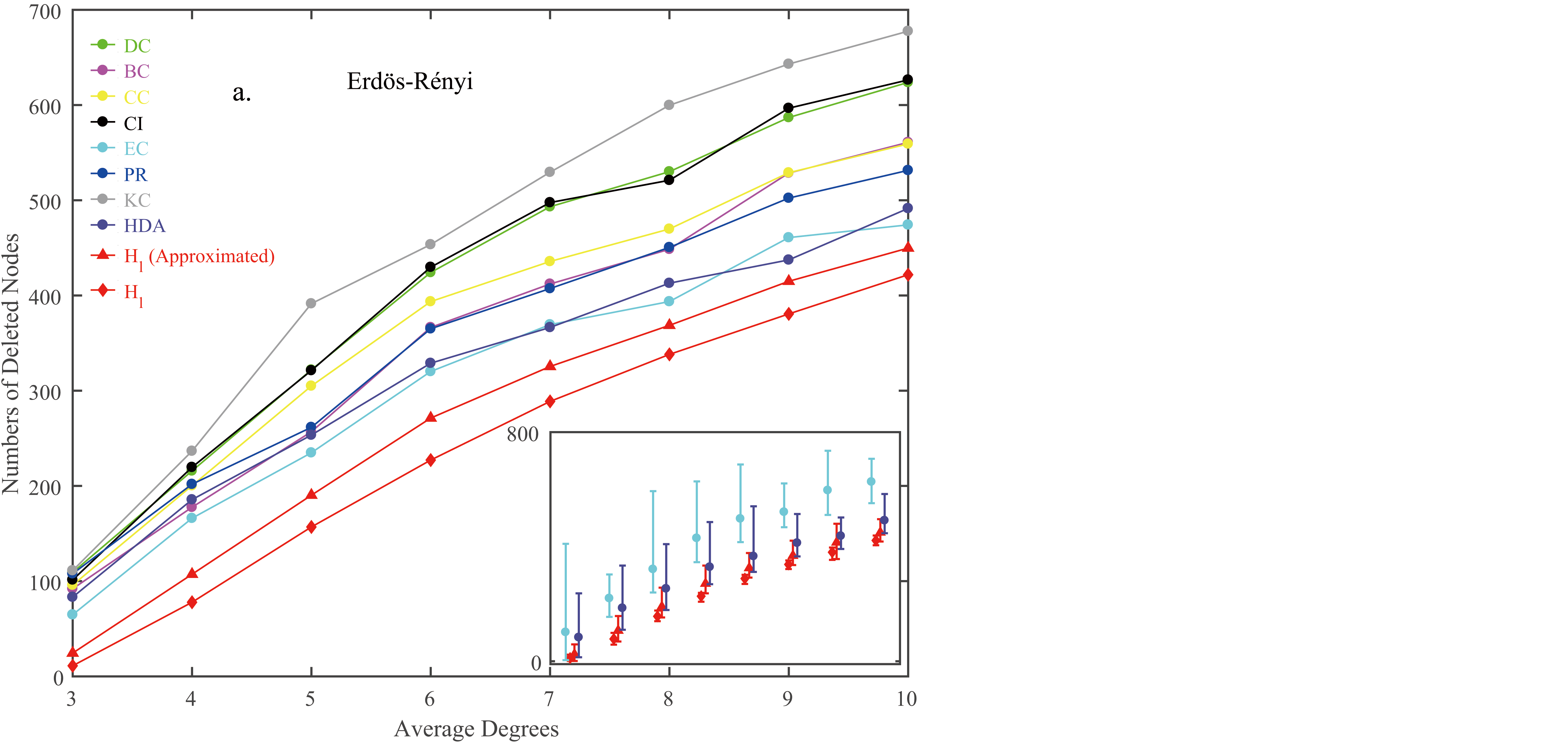}
	\includegraphics[scale=0.07]{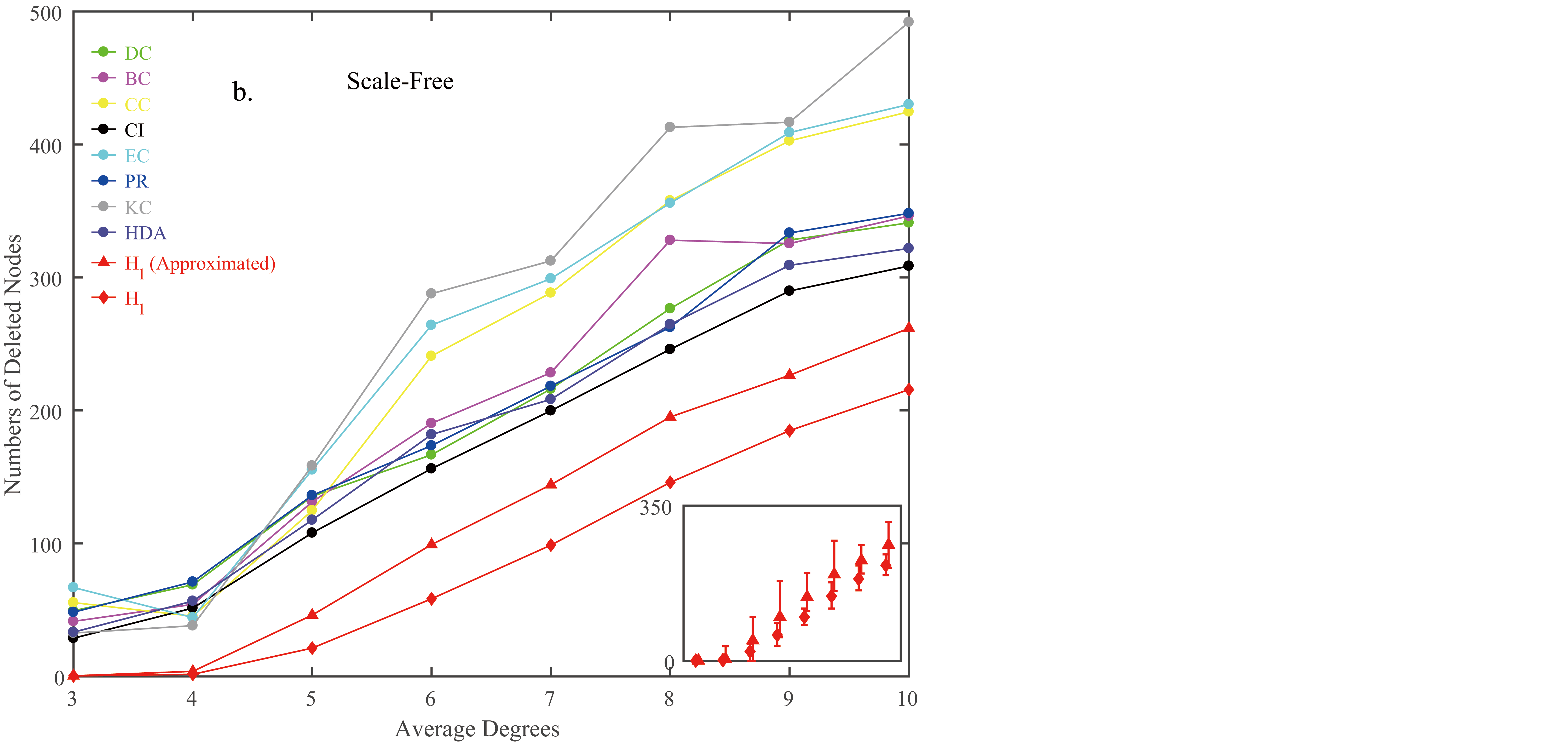}
	\caption{The average numbers of nodes required to be deleted to turn the graphs into no-core ones. The experiments are conducted on ER and SF graphs with average degrees ranging from $3$ to $10$. For each average degree value, $30$ graphs are generated and each graph contains $1000$ nodes. Results of DC, KC, BC, CC, CI, HDA, PR, EC and Core Influence $H_l$ are drawn on the graph. \textbf{a.} Results on ER graphs. The error bars of EC, HDA, $H_l$ and approximated $H_l$ (The four best methods) are plotted in the inside figure. \textbf{b.} Results on Scale-Free graphs with $\gamma=3$.}\label{NF}
\end{figure}

We compare our Core Influence method $H_l$ with centralities including High Degree Centrality (DC) \cite{newman2018networks}, $k$-Core (KC) \cite{dorogovtsev2006kcore}, Betweenness Centrality (BC) \cite{freeman1977set}, Closeness Centrality (CC) \cite{bavelas1950communication, Sabidussi1966}, Collective Influence (CI) \cite{morone2015influence}, High Degree Adaptive (HDA), PageRank (PR) \cite{page1999pagerank} and Eigenvector Centrality (EC) \cite{freeman1978centrality} on Erd\"os-R\'enyi (ER) graphs and Scale-Free (SF) graphs \cite{newman2000networks}. All these methods are applied to calculate how many nodes are required to be removed to turn the random graphs into no-core ones. Results are shown in Figure \ref{NF}.

The Core Influence method is applied at order $l = 1$ in equation~\eqref{Hvi}. As we could see, the Core Influence method performs great efficiency in this problem. With the average degree increasing, the graphs get more denser, yet compared to other node importance indices the Core Influence method still keeps requiring fewer nodes to be deleted to turn the graphs into no-core ones. The Core Influence method with an approximated states updating method in Algorithm~\ref{algo2} performs similar results. Results on SF graphs are better than that on the ER graphs, namely on SF graphs fewer nodes are required to be deleted to transform the graphs into no-core ones. That's because there exist nodes with very high degrees or hubs and deleting them will bring large changes on the topological structure. As a localized greedy method aiming at breaking the Leaf-Removal cores, the proposed Core Influence measurement works more efficiently on different topologies 

\subsection{Results for Minimal Vertex-Cover Number}
\begin{figure}[htb]
	\includegraphics[scale=0.07]{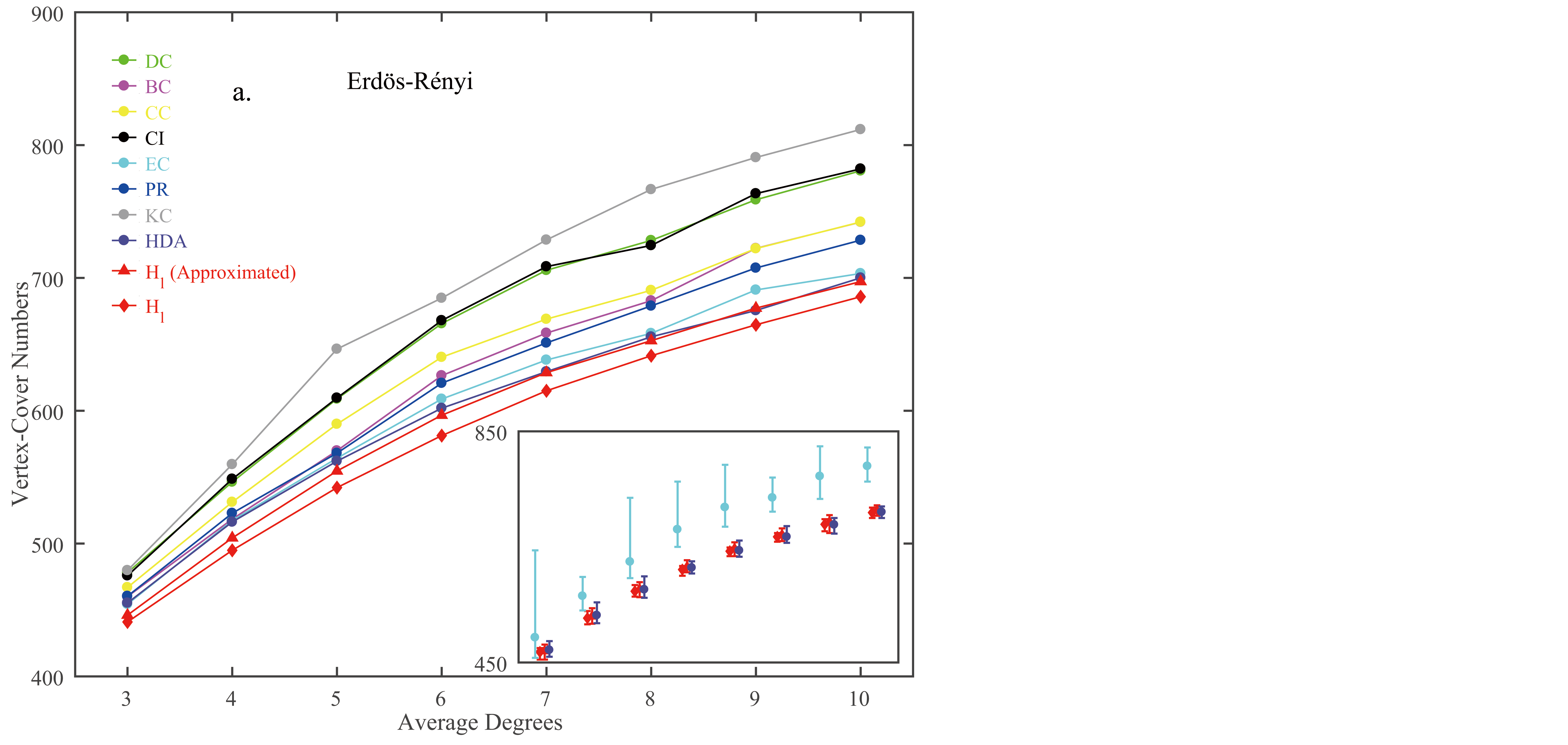}
	\includegraphics[scale=0.07]{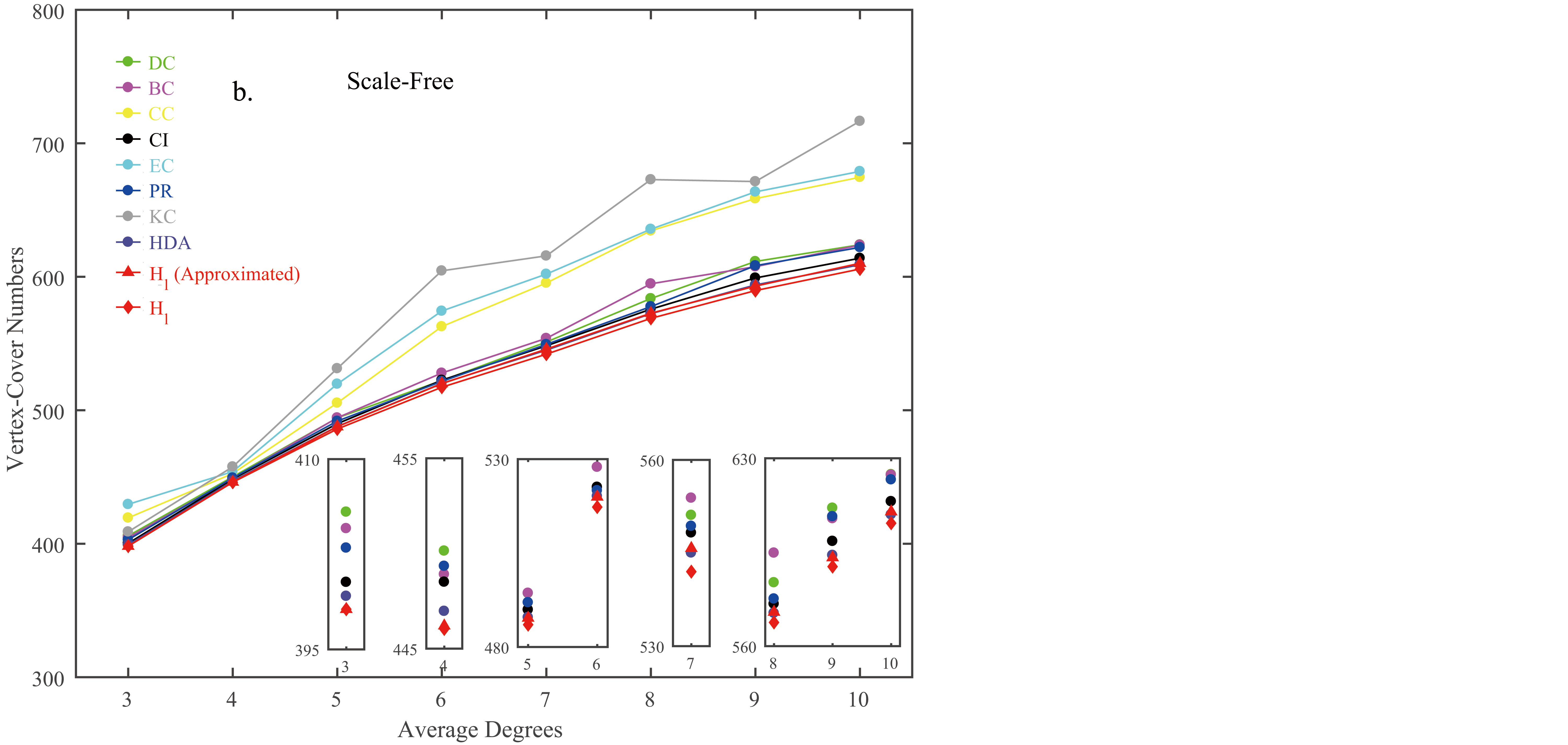}
	\caption{The vertex-cover number by deleting the graphs into no-core ones. The experiments are conducted on ER and SF graphs with average degrees ranging from $3$ to $10$. For each average degree value, $30$ graphs are generated and each graph contains $1000$ nodes. Results of DC, KC, BC, CC, CI, HDA, PR, EC and Core Influence $H_l$ are drawn on the graph. \textbf{a.} Minimal vertex-cover numbers on ER graphs. The error bars of EC, HDA, $H_l$ and approximated $H_l$ (The four best methods) are plotted in the inside figure. \textbf{b.} Minimal vertex-cover numbers on SF graphs. Inside figures are the local amplifications.}\label{CoverNum}
\end{figure}

At the same time, the vertex-cover is determined by the deleted nodes and the remained no-core graphs, and different node importance measurements are applied to delete the graphs into no-core ones and then by equation~\eqref{mvc1} and~\eqref{mvc2} the minimal vertex-cover numbers can be calculated. The specific process for getting the minimal vertex-cover by Core Influence is presented in Algorithm \ref{algo3}. Results of the experiments are shown in Figure \ref{CoverNum}. 

\begin{algorithm}[H]
	\caption{The vertex-cover number by Core Influence}
	\begin{algorithmic}
		\State {\bf Input} graph $G$;
		\State Get the deleted node array $\mathbf{n}$ by Algorithm~\ref{algo1};
		\State Get the maximal matching $M(G(\mathbf{n}))$ for graph $G(\mathbf{n})$;
		\State {\bf Output} The vertex-cover number $N-N(\mathbf{n}) + M(G(\mathbf{n}))$.
	\end{algorithmic}
	\label{algo3}
\end{algorithm}

As we could see, by the Leaf-Removal process, the coverage numbers are much smaller compared to other methods, especially when the average degrees are small. The Core Influence method proposed in the paper works better and could lower thel vertex-covering numbers. When the degree is getting higher (than $6$), the approximated method in Algorithm~\ref{algo2} works similar to the HDA method and they both work better than other centralities, yet the complete algorithm $H_l$ still works better than HDA. These results perform that the solution by deleting the graphs into no-core ones is a new perspective in studying graph-related problems and has shown great potential in minimal vertex-covering problem. More researches on the updating strategy are expected to improve the performance.

\begin{table}[htbp]
	\caption{\label{tab:1} The percentages of differences between results of Core Influence method and exact minimal vertex-cover numbers against nodes numbers. The experiments are conducted on the ER graphs with $80$, $100$ and $120$ nodes with average degrees from $3$ to $7$ and every results are the average of $30$ graphs.}
	\begin{ruledtabular}
		\begin{tabular}{cccccc}
			\multirow{2}{*}{Nodes Numbers} & \multicolumn{5}{c}{Average Degrees}\\
			\cline{2-6} &3&4&5&6&7\\
			\colrule
			$N=80$ & 0.21\% & 0.71\% & 1.33\% & 1.63\% & 1.38\%\\
			$N=100$ & 0.27\% & 0.80\% & 1.53\% & 1.43\% & 1.47\%\\
			$N=120$&0.05\% & 0.67\%& 1.28\% &1.33\% & 1.53\% \\
		\end{tabular}
	\end{ruledtabular}
\end{table}
To further explore the efficiency of the Core Influence method, the cover numbers resulting from this method are compared with the exact results of minimal vertex-cover. The differences between the results of Core Influence method and real $C_m$ are calculated and the percentages of these differences against the total nodes numbers are shown in Table~\ref{tab:1}. As we could see, the Core Influence performs well and as the nodes numbers increase, this method stays stable. With constant nodes number, as the edges numbers increase, the gaps are getting larger. That's due to the increase of the densely connected clusters. Yet the Core Influence method still keeps high efficiency and the cover numbers keep a low difference from the real minimal vertex-cover numbers.

\subsection{Experiments on GR-QC Collaboration Network}

\begin{table*}[htbp]
	\caption{\label{tab:2} The numbers of nodes required to be deleted to turn the General Relativity and Quantum Cosmology (GR-QC) collaboration network into no-core one and the corresponding vertex-cover number. There are $5242$ nodes and $14496$ edges in the network and results of DC, KC, BC, CC, CI, HDA, PR, EC and Core Influence $H_l$ are shown.}
	\begin{ruledtabular}
		\begin{tabular}{ccccccccccc}
			& DC & BC & CC &CI&EC&PR&KC&HDA&$H_l$ & $H_l$(Approximated)\\
			\colrule
			Deleted Nodes Numbers&4042&	5240&	4778&	1563&	5211&	4577&	3918&  1831&	\textbf{779} &	\textbf{855}\\
			Vertex-Cover Numbers & 4220&	5241&	4956&	2820&	5220&	4582&	4213&	2795&	\textbf{2785}&	\textbf{2787}\\
		\end{tabular}
	\end{ruledtabular}
\end{table*}
The same experiments are conducted on the General Relativity and Quantum Cosmology (GR-QC) collaboration network \cite{leskovec2007graph}. This network is from the \emph{arXiv} site and records the papers submitted to the GR-QC category from January 1993 to April 2003. There are $5242$ nodes and $14496$ edges in the network. Each node represents an author and if two authors cooperated and submitted a paper to the GR-QC category in \emph{arXiv} together, there will be an undirected link between the corresponding nodes.

The Leaf-Removal process and Core Influence method are implemented on the network and results are shown in Table~\ref{tab:2}. As the results show, the Core Influence method keeps the great efficiency and require removing less than $1000$ nodes to get the no-core graph. The BC require $5240$ nodes to be deleted to turn the graph into no-core one, which is only $2$ nodes less than the total nodes number. This is due to the appearance of isolated triangles in the Leaf-Removal process, which is common in scientific research cooperations. At the same time, the vertex-cover number by Core Influence method is also the lowest, around half of the total number of nodes. It is hard to get the vertex-cover for networks at this scale, yet we could see the $H_l$ gets reasonable results on the minimal vertex-cover problem.

\section*{Conclusion and Discussion}
In this paper, focusing on the minimal vertex-cover problem, a method based on dynamical system and power method is proposed. This Core Influence method could transform the graphs into No-Leaf-Removal-Core ones fast by deleting selected nodes and breaking the Leaf-Removal cores. It is proved that any minimal vertex-covers of the whole graph can be located into Leaf-Removal Cores breaking and maximal matching of the remained graphs and the best boundary is achieved at the transition point. The coverage numbers resulting from this Core Influence method is better compared to other node importance indices. Since the Leaf-Removal process is crucial in many researches, this study provides a useful tool for researches on graphs and a new perspective of the minimal vertex-cover problem.

Since the structure of Leaf-Removal cores is sophisticated and we still have not made it clear, researches related to it are expected to further improve this method. More understandings on the transition to no-core graphs will help find more accurate expressions on the nodes in Leaf-Removal cores and improve the performance. Also, a fast and accurate method for judging whether a node belongs to the Leaf-Removal cores will improve the accuracy and efficiency. More researches on NP problem in the view of Leaf-Removal process are expected in the future.

\section*{Acknowledgements}
This work is supported by the Fundamental Research Funds for the Central Universities, the National Natural Science Foundation of China (No.11201019), the International Cooperation Project No.2010DFR00700 and Fundamental Research of Civil Aircraft No.MJ-F-2012-04.

\end{document}